\documentclass[12pt]{article}
\usepackage{amssymb}
\usepackage{amsmath}
\usepackage{amscd}
\usepackage{latexsym}
\usepackage{dsfont}
\def\a{\alpha}
\def\b{\beta}

\def\d{\delta}

\def\g{\gamma}

\def\la{\lambda}

\def\be{\begin{equation}}
\def\ee{\end{equation}}
\def\arr{\begin{array}{rll}}
\def\ea{\end{array}}
\def\bea{\begin{eqnarray}}
\def\eea{\end{eqnarray}}

\def\N2{$N{=}2$}

\def\&gt;{\rangle}
\def\&lt;{\langle}
\def\+{\dagger}
\def\={\ =\ }

\begin{document}
\renewcommand{\thefootnote}{\fnsymbol{footnote}}
\begin{titlepage}
\setcounter{page}{0}
\begin{flushright}
LMP-TPU-05/13  \\
\end{flushright}
\vskip 1cm
\begin{center}
{\LARGE\bf Dynamical realizations of $l$-conformal}
\vskip 0.3 cm
{\LARGE\bf  Newton-Hooke group}\\
\vskip 2cm
$
\textrm{\Large Anton Galajinsky, Ivan Masterov\ }
$
\vskip 0.7cm
{\it
Laboratory of Mathematical Physics, Tomsk Polytechnic University, \\
634050 Tomsk, Lenin Ave. 30, Russian Federation} \\
{Emails: galajin@tpu.ru, masterov@tpu.ru}

\end{center}
\vskip 1cm
\begin{abstract} \noindent
The method of nonlinear realizations and the technique previously developed in [Nucl. Phys. B 866 (2013) 212]
are used to construct a dynamical system without higher derivative terms, which holds invariant
under the $l$--conformal Newton-Hooke group. A configuration space of the model involves coordinates, which
parametrize a particle moving in $d$ spatial dimensions and a conformal mode, which gives rise to an effective external field.
The dynamical system describes a generalized multi--dimensional oscillator, which undergoes accelerated/decelerated motion
in an ellipse in accord with evolution of the conformal mode. Higher derivative formulations are discussed as well. It is demonstrated that the multi--dimensional Pais--Uhlenbeck oscillator enjoys the $l=\frac{3}{2}$-conformal Newton-Hooke symmetry for a particular choice of its frequencies.
\end{abstract}


\vskip 0.5cm


\end{titlepage}

\renewcommand{\thefootnote}{\arabic{footnote}}
\setcounter{footnote}0

\noindent
{\bf 1. Introduction}

\vskip 0.7cm

The Newton--Hooke algebra is an analogue of the Galilei algebra in the presence of a universal cosmological repulsion or attraction \cite{bac,gao,gp}.
It is derived from the (anti) de Sitter algebra by the nonrelativistic contraction in the same way as
the Galilei algebra is obtained from the Poincar\'e algebra \cite{bac}. A peculiar feature of the Newton--Hooke algebra is that
the structure relation involving the generators of
time and space translations yields the Galilei boost: $[H,P_i]=\pm \frac{1}{R^2} K_i$. The positive constant $R$ is called the characteristic time.\footnote{Being multiplied with the speed of light, it yields
the radius of the parent (anti) de Sitter space.} The upper/lower sign above is realized in nonrelativistic spacetime with the negative/positive cosmological constant $\Lambda=\mp \frac{1}{R^2}$. In the flat space limit
$\Lambda \to 0$ the Galilei algebra is recovered.

Conformal extensions of the Galilei algebra have recently attracted considerable interest mostly due to potential uses within the context of the nonrelativistic $AdS/CFT$--correspondence.
Such extensions are parameterized by a positive half--integer $l$ \cite{Havas,nor}, thus justifying the term the $l$--conformal Galilei algebra \cite{nor}.\footnote{In modern literature the reciprocal $N=\frac{1}{l}$ is called the rational dynamical exponent.
The corresponding
algebra is referred to as
the conformal Galilei algebra with rational dynamical exponent or $N$--Galilean conformal algebra (see e.g. \cite{Horvathy_1,Gomis}). In this work we use the terminology
originally introduced in \cite{nor}.}
In general, $(2l+1)$ vector generators enter the algebra, which along with spatial
translations and Galilei boosts involve accelerations. In constructing dynamical realizations, the generators in the algebra are linked to constants of the motion.
Because the number of functionally independent constants of the motion needed to integrate
a differential equation correlates with its order, dynamical realizations of the $l$--conformal Galilei algebra in general involve higher
derivative terms (see e.g. \cite{Horvathy_1}--\cite{Andrzejewski_1}). Dynamical realizations without higher derivatives have been constructed recently in
\cite{Fedoruc,Galajinsky_1} within the method of nonlinear realizations. In particular, it was demonstrated in \cite{Galajinsky_1} that, while
the accelerations are involved in the formal algebraic structure behind the generalized oscillator equations in \cite{Fedoruc,Galajinsky_1},
they prove to be functionally dependent.

Although the $l$--conformal extension of the Newton--Hooke algebra is known for a long time \cite{nor1,Galajinsky_2} \footnote{Note that the flat space limit of the $l$--conformal Newton--Hooke algebra in \cite{nor1} does not yield the $l$--conformal Galilei algebra. This shortcoming was overcame in \cite{Galajinsky_2}.}, its dynamical realizations remain almost completely unexplored (for related earlier studies see \cite{Liu}--\cite{papadop}).
In this work, we continue the analysis initiated in \cite{Galajinsky_1} and construct dynamical realizations of the $l$--conformal Newton--Hooke group in terms of second order differential equations.
Higher derivative formulations are considered as well. In particular, it is shown that the multi--dimensional Pais--Uhlenbeck oscillator enjoys the $l=\frac{3}{2}$--conformal
Newton-Hooke symmetry for a particular choice of its frequencies.

The work is organized as follows. In Sect. 2 the method of nonlinear realizations and the technique previously developed in \cite{Galajinsky_1} are used to construct a
dynamical realization without higher derivatives for the $l$--conformal
Newton-Hooke group with $\Lambda<0$. Similar analysis for $\Lambda>0$ is presented in Sect. 3.
Higher derivative formulations are discussed in Sect. 4. In particular, it is shown that a variant of the multi--dimensional Pais--Uhlenbeck oscillator is invariant under
the $l=\frac{3}{2}$--conformal Newton-Hooke symmetry. In Sect. 5 we summarize our results and discuss possible further developments.

\newpage

\noindent
{\bf 2.  Dynamical realization of $l$-conformal Newton-Hooke algebra with $\Lambda<0$}

\vskip 0.5cm
The $l$-conformal Newton-Hooke algebra involves the generator of time translations $H$, the generator of dilatations $D$, the generator
of special conformal transformations $K$, the set of vector generators $C_i^{(n)}$ with $n=0,\dots,2l$, $i=1,\dots,d$, and the generators of spatial rotations $M_{ij}$. The structure relations for the algebra read \cite{Galajinsky_2}
\bea\label{NH}
&&
[H,D]=i\left(H\mp\frac{2}{R^2}K\right),\qquad\quad [H,C_i^{(n)}]=i\left(nC_i^{(n-1)}\pm\frac{(n-2l)}{R^2}C_i^{(n+1)}\right),
\\[2pt]
&&
[H,K]=2iD,\qquad\qquad\qquad\qquad [D,K]=iK,
\nonumber\\[2pt]
&&
[D,C_i^{(n)}]=i(n-l)C_i^{(n)},\qquad\quad\;\;\, [K,C_i^{(n)}]=i(n-2l)C_i^{(n+1)},
\nonumber\\[2pt]
&&
[M_{ij},C_k^{(n)}]=i\d_{jk}C_i^{(n)}-i\d_{ik} C_j^{(n)},\; [M_{ij},M_{kl}]=i\left(\d_{jk}M_{il}+\d_{il}M_{jk}-\d_{ik}M_{jl}-\d_{jl}M_{ik}\right).
\nonumber
\eea
In the first line the upper/lower sign corresponds to a negative/positive cosmological constant. Below we construct a dynamical realization of the
$l$-conformal Newton-Hooke algebra in terms of second order differential equations for the case of a negative cosmological constant.
Positive cosmological constant is discussed in Sect. 3.

In order to construct a dynamical realization, we resort to the method of nonlinear realizations \cite{Coleman_1}--\cite{Ivanov_1}. The first ingredient is the coset space
arising from the quotient by a subgroup of rotations\footnote{Here and in what follows the summation over repeated indices is understood.}
\bea\label{coset}
G=e^{itH}e^{izK}e^{iuD}e^{ix_i^{(n)}C_i^{(n)}},
\eea
which is
parametrized by the real coordinates $t$, $u$, $z$ and $x_i^{(n)}$. Left multiplication by the group element $g=e^{iaH}e^{ibK}e^{icD}e^{i\la_i^{(n)}C_i^{(n)}} e^{\frac{i}{2} \omega_{ij} M_{ij}}$ determines the action of the group on the coset.
For practical applications the infinitesimal form is enough. In particular, for the case at hand one finds the following generators of the infinitesimal transformations\footnote{As usual, given a generator $\tilde T$, an infinitesimal parameter $\alpha$ and a coordinate $y$, the infinitesimal transformation reads $y'=y+\alpha {\tilde T} y$, where ${\tilde T}$ is to be understood as a differential operator acting on $y$.}
\bea\label{gen}
&&
\tilde H=\frac{\partial}{\partial t},\qquad
\tilde D=\frac{R}{2}\sin{\frac{2t}{R}}\frac{\partial}{\partial t}+\cos{\frac{2t}{R}}\frac{\partial}{\partial u}-\left(\frac{1}{R}\sin{\frac{2t}{R}}+z\cos{\frac{2t}{R}}\right)\frac{\partial}{\partial z},
\\[2pt]
&&
\tilde K=\frac{R^2}{2} \left(1- \cos{\frac{2t}{R}}\right) \frac{\partial}{\partial t}+R\sin{\frac{2t}{R}}\frac{\partial}{\partial u}+\left(\cos{\frac{2t}{R}}-zR\sin{\frac{2t}{R}}\right)\frac{\partial}{\partial z},
\nonumber\\[2pt]
&&
{\tilde C}_i^{(0)}=\sum_{n=0}^{2l}\sum_{m=0}^n \frac{(-1)^n}{R^m}\frac{(2l)!}{m!(n-m)!(2l-n)!}\left(\cos{\frac{t}{R}}\right)^{2l-m}\left(\sin{\frac{t}{R}}\right)^m z^{n-m}e^{u(n-l)}\frac{\partial}{\partial x_i^{(n)}},
\nonumber\\[2pt]
&&
{\tilde M}_{ij}= x_i^{(n)} \frac{\partial}{\partial x_j^{(n)}}-x_j^{(n)} \frac{\partial}{\partial x_i^{(n)}},
\nonumber
\eea
while ${\tilde C}_i^{(n)}$ with $n>0$ are obtained from $C_i^{(0)}$ and $K$ by computing the successive commutators
\be
{\tilde C}_i^{(n)}=\frac{(-1)^n (2l-n)!}{(2l)!}\underbrace{[\tilde K,[ \tilde K,..[\tilde K,{\tilde C}_i^{(0)}]..]]}_{n\,times}.
\ee
In obtaining (\ref{gen}) the standard Baker-Campbell-Hausdorf formula proves to be helpful.
Then one constructs the Maurer-Cartan one-forms
\be
G^{-1}dG=i\left(w_H H+w_D D+w_K K+w_i^{(n)} C_i^{(n)}\right),
\ee
where
\bea\label{MC1}
&&
w_H=e^{-u}dt, \qquad w_D=du-2z dt, \qquad w_K=e^u \left( dz+z^2 dt \right)+\frac{2}{R^2} \sinh{u} dt,
\\[2pt]
&&
w_i^{(n)}=dx_i^{(n)}-(n+1) x_i^{(n+1)} w_H-(n-l)x_i^{(n)}w_D-(n-2l-1)x_i^{(n-1)} \left(w_K+\frac{1}{R^2} w_H\right),
\nonumber
\eea
which are invariant under all the transformations from the $l$--conformal Newton-Hooke group but for rotations with respect to which $w_i^{(n)}$ is transformed as a vector.
It is assumed in (\ref{MC1}) that $x^{(-1)}_i=x^{(2l+1)}_i=0$.

The Maurer-Cartan one-forms are the building blocks of a dynamical realization. Setting some of them to zero
one can either reduce the number of degrees of freedom via algebraic equations or obtain dynamical equations of motion, which are automatically
invariant under the action of a given group \cite{Ivanov_1}. In this work we choose the following constraints
\bea\label{Constr}
w_D=0,\qquad \tilde{\g}^{-1}w_K-\tilde{\g} w_H=0, \qquad w_i^{(n)}=0,
\eea
where $\tilde{\g}$ is an arbitrary (coupling) constant. The first two restrictions in (\ref{Constr}) are the conventional constraints associated with the conformal subalgebra $so(2,1)$ \cite{Ivanov_2}.
Taking $t$ to be the temporal coordinate and introducing the new variable
\bea\label{rho}
\rho=e^{\frac{u}{2}}
\eea
from the first two equations in (\ref{Constr}) one readily gets
\bea\label{Equ1}
z=\frac{\dot{\rho}}{\rho}, \qquad \ddot{\rho}=\frac{\gamma^2}{\rho^3}-\frac{\rho}{R^2},
\eea
where we abbreviated $\g^2=\tilde{\g}^2+\frac{1}{R^2}$ and the dot denotes the derivative with respect to time.
From (\ref{Equ1}) one concludes that $z$ is not independent and can be discarded, while $\rho$ describes the conformal particle in one
dimension in the harmonic trap. This degree of freedom is conveniently described by the effective action functional
\bea\label{action}
S_{\rho}=\int dt\left(\dot{\rho}^2-\frac{\g^2}{\rho^2}-\frac{\rho^2}{R^2}\right).
\eea
In particular, applying the Noether theorem to the infinitesimal conformal symmetry transformations, which can be deduced from (\ref{gen}), one gets the conserved
charges
\bea\label{const}
&&
\mathcal{H}=\dot{\rho}^2+\frac{\g^2}{\rho^2}+\frac{\rho^2}{R^2}, \qquad \mathcal{D}=\rho\dot{\rho} \cos{\frac{2t}{R}}-\frac{1}{2} \mathcal{H} R\sin{\frac{2t}{R}}+\frac{1}{R} \rho^2 \sin{\frac{2t}{R}},
\nonumber\\[2pt]
&&
\mathcal{K}=\rho^2 \cos{\frac{2t}{R}}- \rho\dot{\rho} R \sin{\frac{2t}{R}}+\frac 12 \mathcal{H} R^2 \left(1- \cos{\frac{2t}{R}} \right) ,
\eea
which allow one to determine the dynamics of $\rho$ by purely algebraic means
\bea\label{solution}
\rho(t)=\sqrt{\frac{\left( \mathcal{D} R\sin{\frac{t}{R}}+\mathcal{K} \cos{\frac{t}{R}}\right)^2+{\left( \g R \sin{\frac{t}{R}} \right)}^2}{\mathcal{K}}}.
\eea
Note that the conserved charges (\ref{const}) are functionally dependent
\be
\mathcal{H}\mathcal{K}-\mathcal{D}^2-\frac{\mathcal{K}^2}{R^2}=\g^2.
\ee
From the latter relation it also follows that $\mathcal{K}$ cannot be zero or negative, which is essential in Eq. (\ref{solution}) above.

Now let us turn to the rightmost constraint in (\ref{Constr}) which determines $x_i^{(n)}$. Taking into account the first two relations in (\ref{Constr}) and Eqs. (\ref{MC1}) and (\ref{rho}), one readily finds
\bea\label{Equ2}
\rho^2 {\dot x}_i^{(n)}=(n+1)x_i^{(n+1)}+(n-2l-1)\g^2 x_i^{(n-1)}.
\eea
Beautifully enough, this system of ordinary differential equations is analogous to that, which appeared recently in \cite{Galajinsky_1} in constructing dynamical realizations for the $l$--conformal Galilei algebra.
There are two essential differences, however. As compared to flat spacetime, the effective frequency of oscillations is increased due to the contribution proportional to the cosmological constant
\be
\tilde{\g}^2 \quad \rightarrow \quad \g^2=\tilde{\g}^2+\frac{1}{R^2}
\ee
and the external field provided by the conformal mode $\rho_G (t)$ \cite{Galajinsky_1}
\be\label{CM}
\rho_G (t)=\sqrt{\frac{{(\mathcal{D}+t\mathcal{H})}^2+\tilde\gamma^2}{\mathcal{H}}},
\ee
is promoted to (\ref{solution}).

The system (\ref{Equ2}) can be treated in two different ways. Taken literally, it is equivalent to one dynamical equation, which involves higher derivative terms \cite{Gomis,Andrzejewski_1,Galajinsky_1}.
On the other hand, it can be rewritten as a collection of subsystems of differential equations \cite{Galajinsky_1}, each of which is invariant under the $l$--conformal Newton--Hooke group and involves at most two derivatives.
In order to construct such invariant subsystems, it proves sufficient to rewrite (\ref{Equ2}) in the matrix form
\bea\label{matrix}
\rho^2 {\dot x}_i^{(n)}=x_i^{(m)}A^{mn}
\eea
and find eigenvalues and eigenvectors of $A^{mn}$. As was demonstrated in \cite{Galajinsky_1}, the eigenvalues appear in complex conjugate pairs $\pm i p \gamma$, where
$p=0,2,4,\dots,2l$ for an integer $l$ and $p=1,3,\dots,2l$ for a half--integer $l$. The corresponding eigenvectors go in pairs as well, which we denote by $v^{p}_{(n)}$ and ${\bar v}^{p}_{(n)}$. The desired dynamical realization is constructed by analogy with \cite{Galajinsky_1}
\bea\label{result}
\ddot{\rho}=\frac{\g^2}{\rho^3}-\frac{\rho}{R^2},\qquad \rho^2\frac{d}{dt}\left(\rho^2\frac{d}{dt}\chi_i^p\right)+(p\g)^2\chi_i^p=0,
\eea
where the dynamical field $\chi_i^p$ is built from the eigenvectors of $A^{mn}$ and the original variables $x^{(n)}_i$ in the following way:
\be\label{def}
\chi_i^p=x^{(n)}_i (v^p_{(n)}+{\bar v}^{p}_{(n)}).
\ee
It turns out that the option $p=0$ leads to a first order differential equation \cite{Galajinsky_1}, which should be discarded on physical grounds.
On the space of fields $\rho(t)$ and $\chi_i^p (t)$ the $l$--conformal Newton-Hooke group acts as follows:
\be
\rho' (t')=\rho(t)+\delta \rho, \qquad {\chi'}_i^p (t')=\chi_i^p(t)+\delta \chi_i^p,
\ee
where $\delta \rho$ and $\delta \chi_i$ are inherited from the coset transformations with the generators (\ref{gen}).

Thus, for any value of $p$ from the range $1,\dots,2l$ one can construct a pair of the second order differential equations (\ref{result}), which provide a dynamical realization of the $l$--conformal Newton-Hooke group.
Focusing on integer $l$, one can notice that the value ${(2 \gamma )}^2$ appears for
any $l$, the equation involving ${(4 \gamma)}^2$ is common for all $l>1$, the frequency ${(6 \gamma)}^2$ is shared by all $l>2$ etc. A similar observation can be made for a half--integer $l$ as well.
The reason why one can realize different $l$--conformal Newton-Hooke groups in one and the same equation is that all the vector generators $C^{(n)}_i$ with
$n>1$ prove to be functionally dependent (see Ref. \cite{Galajinsky_1} for more details).
Although $C^{(n)}_i$ with $n>1$ are involved in the
formal algebraic structure leading to (\ref{result}),
they prove to be irrelevant for actual solving thereof.

Because the variables $\rho$ and $\chi_i^p$ are separated in (\ref{result}), one can solve the differential equation for the conformal mode $\rho$ and substitute its solution (\ref{solution}) to the equation for $\chi_i^p$.
The latter then describes a particle in $d$ spatial dimensions moving in an external field. The general solution of the rightmost equation in (\ref{result}) is readily obtained by introducing a new variable $s$ such that
\be\label{s}
\rho^2 \frac{d}{dt}=\frac{d}{ds} \quad \rightarrow \quad  \frac{ds}{dt}=\frac{1}{\rho^2} \quad \rightarrow \quad  s(t)=\frac{1}{\g}\arctan{\frac{\mathcal{D}\mathcal{K}+(\mathcal{D}^2+\g^2) R\tan{\frac{t}{R}}}{\g\mathcal{K}}}.
\ee
This yields
\be\label{CHI}
\chi_i^p=\alpha_i^p \cos{(p\g s(t))}+\beta_i^{p}\sin{(p\g s(t))},
\ee
where $\alpha_i^p$ and $\beta_i^{p}$ are constants of integration. We assume $\gamma$ to be positive.

Let us confront a qualitative behavior of a particle described by Eq. (\ref{CHI}) with that invariant under the $l$--conformal Galilei symmetry. In the latter case, the solution reads as in  (\ref{CHI}),
while the subsidiary function $s(t)$ takes the form \cite{Galajinsky_1}
\be\label{subs}
s(t)=\frac{1}{\tilde\gamma} \arctan{\left(\frac{\mathcal{D}+t \mathcal{H}}{\tilde\gamma}\right)} \quad \rightarrow \quad \dot s(t)=\frac{1}{\rho_G^2 (t)}
\ee
with $\rho_G (t)$ displayed above in Eq. (\ref{CM}). Here $\mathcal{D}$ and $\mathcal{H}$ are constants of the motion and $\tilde\gamma$ is the coupling constant corresponding to the conventional conformal mechanics in $d=1$ without the oscillator potential. We assume $\tilde\gamma$ to be positive.
For definiteness, in what follows we choose $p=2$ ($l=1$) and stick to the case of three spatial dimensions. Making use of the rotation invariance, one can choose a coordinate system in which the motion occurs in the $xy$--plane, with $\alpha_i$ \footnote{Here and in what follows we omit the superscript $p=2$ attached to $\alpha_i$ and $\chi_i$.} in (\ref{CHI}) being parallel to the $x$--axis. The orbit is an ellipse with one point corresponding to the polar angle $\phi=2\tilde\gamma s=\pi$ removed. One can readily verify that, being initially at rest close to $\chi_i=-\alpha_i$ (as $t\to -\infty$), the particle starts moving towards $\chi_i=\alpha_i$ with growing angular velocity $\frac{d \phi}{dt}=2\tilde\gamma\frac{d s}{dt}=\frac{2 \tilde\gamma}{\rho_G^2 (t)}$. Given the initial data $\mathcal{D}$, $\mathcal{H}$, it arrives there at $t=-\frac{\mathcal{D}}{\mathcal{H}}$, which corresponds to the polar angle $\phi=2\tilde\gamma s=0$. Then it continues to move towards $\chi_i=-\alpha_i$ with decreasing angular velocity and freezes up as $t\to \infty$.

For a particle invariant under the $l$--conformal Newton-Hooke group the shape of the orbit is the same but a qualitative behavior is different. The range of the temporal coordinate which enters Eqs. (\ref{s}) and (\ref{CHI}) is $-\frac{\pi R}{2}<t<\frac{\pi R}{2}$.
As $t\to \pm \frac{\pi R}{2}$ the angular velocity $\frac{d \phi}{dt}=2 \gamma\frac{d s}{dt}=\frac{2 \gamma}{\rho^2}$ tends to the constant value
$\left(\frac{2 \gamma \mathcal{K}}{\mathcal{D}^2+\gamma^2} \right) \frac{1}{R^2}$, which is proportional to the cosmological constant $\Lambda=\frac{1}{R^2}$.
If $\mathcal{D}>0$, one reveals three regimes in which the angular velocity first increases, then decreases and then increases again. Likewise, for $\mathcal{D}<0$
two phases of decelerated motion are separated by the acceleration phase in the middle.
In both the cases the three regimes are separated by two roots of $\tan{\frac{2 t}{R} }=\frac{2 \mathcal{D} \mathcal{K} R }{\mathcal{K}^2-(\mathcal{D}^2+\gamma^2) R^2}$.
For $\mathcal{D}=0$ there are two regimes, in which acceleration is followed by deceleration for $\gamma^2-\frac{\mathcal{K}^2}{R^2}>0$ and vice versa for $\gamma^2-\frac{\mathcal{K}^2}{R^2}<0$ . If $\mathcal{D}=0$ and $\mathcal{K}=\gamma R$ the motion is uniform.
Note that, because the angular velocity is fully determined by the conformal mode, the qualitative difference in the motion of a particle along the orbit in the case of the $l$--conformal Galilei symmetry and its Newton-Hooke counterpart correlates with the dynamics of $\rho(t)$. For the former case, the conformal mode is scattered off the repulsive potential $\frac{\tilde\gamma^2}{\rho^2}$ and its motion is unbounded, while for the latter case it is confined to move in the potential well  $\frac{\g^2}{\rho^2}+\frac{\rho^2}{R^2}$. One can verify that for $p>2$ the qualitative picture is similar but a particle makes more than one revolution in the ellipse.

\vskip 0.5cm

\noindent
{\bf 3. Dynamical realization of $l$-conformal Newton-Hooke algebra for $\Lambda>0$}

\vskip 0.5cm

For a positive cosmological constant the construction of a dynamical realization proceeds along similar lines. In particular, choosing the same constraints as in (\ref{Constr}), one gets the following equation of motion for the conformal mode:
\bea\label{rho+}
\ddot{\rho}=\left(\tilde{\g}^2-\frac{1}{R^2}\right)\frac{1}{\rho^3}+\frac{\rho}{R^2}.
\eea
As compared to the previous case, the full potential $\left(\tilde{\g}^2-\frac{1}{R^2}\right)\frac{1}{\rho^2}-\frac{\rho^2}{R^2}$ is unbounded below because of the flip of sign in the oscillator contribution.
This leads to the well known difficulties in treating $\rho(t)$ as a stable dynamical system. Yet, regarding $\rho(t)$ as a fixed background field, one can construct a reasonable dynamical equation for
$x_i^{(n)} (t)$, which is subject to the rightmost constraint in (\ref{Constr}).

The analysis depends on whether $\left(\tilde{\g}^2-\frac{1}{R^2}\right)$ is positive, negative or zero. If the constant is positive, the general solution of (\ref{rho+}) reads
\bea\label{solution+}
\rho(t)=\sqrt{\frac{\left(\mathcal{D} R\sinh{\frac{t}{R}}+ \mathcal{K}\cosh{\frac{t}{R}}\right)^2+{\left( \g R \sinh{\frac{t}{R}}\right)}^2 }{\mathcal{K}}},
\eea
where we abbreviated $\gamma^2=\left(\tilde{\g}^2-\frac{1}{R^2}\right)$. Applying the method in \cite{Galajinsky_1}, one then derives the dynamical equation for the vector variables
\bea\label{vec}
\rho^2\frac{d}{dt}\left(\rho^2\frac{d}{dt}\chi_i^p\right)+(p\g)^2\chi_i^p=0,
\eea
which is the same as the rightmost equation in (\ref{result}) but for the explicit form of the background field $\rho(t)$. As above, the fields $\chi_i^p$ are constructed form the eigenvectors of the matrix (\ref{matrix}) in accord with (\ref{def}) and the range of $p$ is $p=2,4,\dots,2l$ for an integer $l$ and $p=1,3,\dots,2l$ for a half--integer $l$. Given $p$, the pair (\ref{rho+}) and (\ref{vec}) thus provides a dynamical realization of the
$l$-conformal Newton-Hooke algebra for the case of a positive cosmological constant.

The general solution of Eq. (\ref{vec}) has the form (\ref{CHI}), in which the subsidiary function $s(t)$ now reads
\bea
s(t)=\frac{1}{\g}\arctan{\frac{\mathcal{D}\mathcal{K}+(\mathcal{D}^2+\g^2)R\tanh{\frac{t}{R}}}{\g\mathcal{K}}}, \qquad \dot s (t)=\frac{1}{\rho^2}.
\eea
For example, choosing $p=2$ and $d=3$, one reveals accelerated motion of a particle in the ellipse, which turns into decelerated motion at $\tanh{\frac{2t}{R}}=-\frac{2 \mathcal{D} \mathcal{K} R}{\mathcal{K}^2+(\mathcal{D}^2+\gamma^2) R^2}$. This again correlates with the evolution of $\rho(t)$.

If $\left(\tilde{\g}^2-\frac{1}{R^2}\right)$ is negative, one abbreviates $\left(\tilde{\g}^2-\frac{1}{R^2}\right)=-\gamma^2$ and obtains the dynamical equations and their solutions by the formal substitution
$\gamma \to i\gamma$ in the equations above. However, because in this case $\rho(t)$ may fall into the center, the velocity of $\chi_i^p$ may grow unbounded and the solution should be discarded as unphysical.

Finally, if $\left(\tilde{\g}^2-\frac{1}{R^2}\right)$ is zero, the method in \cite{Galajinsky_1} fails to produce second order differential equations and the resulting dynamical realization involves higher derivative terms.

\vskip 0.5cm

\noindent
{\bf 4.  $l=\frac{3}{2}$ conformal Newton-Hooke symmetry in Pais--Uhlenbeck oscillator}

\vskip 0.5cm

In two preceding sections we were mainly concerned with the construction of dynamical realizations for the $l$-conformal Newton-Hooke group in terms of second order differential equations. Yet,
under certain circumstances, higher derivative formulations may turn out to be useful as well. Below we demonstrate that the multi--dimensional Pais--Uhlenbeck oscillator enjoys
the $l=\frac{3}{2}$ conformal Newton-Hooke symmetry for a special choice of its frequencies.

Let us consider the case of a negative cosmological constant and stick to $l=\frac{3}{2}$. The dynamical equations (\ref{Equ2}) for the vector coordinates read
\begin{align}\label{l32}
&
\rho^2 \dot x^{(0)}_i=x^{(1)}_i, && \rho^2 \dot x^{(1)}_i=2 x^{(2)}_i-3 \gamma^2 x^{(0)}_i,
\nonumber\\[2pt]
&
\rho^2 \dot x^{(2)}_i=3 x^{(3)}_i-2 \gamma^2 x^{(1)}_i, && \rho^2 \dot x^{(3)}_i=-\gamma^2 x^{(2)}_i.
\end{align}
Instead of rewriting them as two second order differential equations invariant under the $l=\frac{3}{2}$ conformal Newton-Hooke group, let us eliminate $x_i^{(1)}$, $x_i^{(2)}$, $x_i^{(3)}$ via the algebraic equations
entering (\ref{l32}), and derive the fourth order differential equation for the remaining  $x_i^{(0)}$ (in what follows we omit the superscript attached to $x_i^{(0)}$)
\bea\label{PUgen}
\rho^2\frac{d}{dt}\left(\rho^2\frac{d}{dt}\left(\rho^2\frac{d}{dt}\left(\rho^2\frac{d}{dt}x_i\right)\right)\right)+10\g^2\rho^2\frac{d}{dt}\left(\rho^2\frac{d}{dt}x_i\right)+9\g^4x_i=0.
\eea
It is assumed that $\rho$ obeys the leftmost equation in (\ref{result}). In particular, choosing the static solution for the conformal mode $\rho=\sqrt{\g R}$, which means that $\rho(t)$ is at rest at
the minimum of the corresponding potential $\frac{\g^2}{\rho^2}+\frac{\rho^2}{R^2}$, one arrives at
\bea\label{PU}
x_i^{(4)}+\frac{10}{R^2} {\ddot x}_i+\frac{9}{R^4}x_i=0.
\eea
This is a particular case of the multi--dimensional Pais--Uhlenbeck oscillator $x_i^{(4)}+\alpha^2 {\ddot x}_i+\beta^2 x_i=0$, which in general involves two arbitrary frequencies $\alpha^2$ and $\beta^2$.

Let us demonstrate that (\ref{PU}) enjoys the $l=\frac{3}{2}$ conformal Newton-Hooke symmetry. The action functional reproducing (\ref{PU}) reads
\bea\label{actionPU}
S=\int \left({\ddot x}_i^2-\frac{10}{R^2}\dot{x}_i^2+\frac{9}{R^4} x_i^2\right)dt.
\eea
Taking into account the explicit form of the vector generators in (\ref{gen}), the relations (\ref{rho}), (\ref{Equ1}) which link $u$ and $z$ to $\rho$, and the explicit form of the static $\rho$ above, one
obtains the following transformation laws
\bea\label{transform}
x_i'(t')=x_i(t)+\la_i^{(0)}\cos^3{\frac{t}{R}}+\la^{(1)}_i R\cos^2{\frac{t}{R}}\sin{\frac{t}{R}}+\la^{(2)}_i R^2\cos{\frac{ t}{R}}\sin^2{\frac{t}{R}}+\la^{(3)}_i R^3\sin^3{\frac{t}{R}},
\eea
which are associated with the generators $C_i^{(0)}$, $C_i^{(1)}$, $C_i^{(2)}$ and $C_i^{(3)}$ in the algebra. The Noether theorem then yields constants of the motion
\bea
&&
\mathcal{C}_i^{(0)}=x_i^{(3)}\cos^3{\frac{t}{R}}+3 {\ddot x}_i \frac{1}{R}\cos^2{\frac{t}{R}}\sin{\frac{t}{R}}+7\dot{x}_i\frac{1}{R^2}\cos^3{\frac{t}{R}}+6\dot{x}_i\frac{1}{R^2}\cos{\frac{t}{R}}\sin^2{\frac{t}{R}}+
\nonumber\\[2pt]
&&
\qquad \quad+9 x_i\frac{1}{R^3}\cos^2{\frac{t}{R}}\sin{\frac{t}{R}}+6 x_i\frac{1}{R^3}\sin^3{\frac{t}{R}},
\nonumber\\[2pt]
&&
\mathcal{C}_i^{(1)}=x_i^{(3)} R\cos^2{\frac{t}{R}}\sin{\frac{t}{R}}+2 {\ddot x}_i \cos{\frac{t}{R}}\sin^2{\frac{t}{R}}-{\ddot x}_i \cos^3{\frac{t}{R}}+3\dot{x}_i\frac{1}{R} \cos^2{\frac{t}{R}}\sin{\frac{t}{R}}+
\nonumber\\[2pt]
&&
\qquad \quad+2\dot{x}_i\frac{1}{R}\sin^3{\frac{t}{R}}-3x_i\frac{1}{R^2} \cos^3{\frac{t}{R}},
\nonumber\\[2pt]
&&
\mathcal{C}_i^{(2)}=R^2 x_i^{(3)}\cos{\frac{t}{R}}\sin^2{\frac{t}{R}}+{\ddot x}_i R\sin^3{\frac{t}{R}}-2 {\ddot x}_i R\cos^2{\frac{t}{R}}\sin{\frac{t}{R}}+3\dot{x}_i\cos{\frac{t}{R}}\sin^2{\frac{t}{R}}+
\nonumber\\[2pt]
&&
\qquad \quad +2\dot{x}_i\cos^3{\frac{t}{R}}+3x_i\frac{1}{R}\sin^3{\frac{t}{R}},
\nonumber\\[2pt]
&&
\mathcal{C}_i^{(3)}=x_i^{(3)} R^3\sin^3{\frac{t}{R}}-3 {\ddot x}_i R^2\sin^2{\frac{t}{R}}\cos{\frac{t}{R}}+7\dot{x}_i R\sin^3{\frac{t}{R}}+6\dot{x}_i R\sin{\frac{t}{R}}\cos^2{\frac{t}{R}}-
\nonumber\\[2pt]
&&
\qquad \quad-6x_i\cos^3{\frac{t}{R}}-9x_i\sin^2{\frac{t}{R}}\cos{\frac{t}{R}}.
\eea
This is a set of linear algebraic equations, which allows one to fix $x_i$ (and its derivatives up to the third order) by purely algebraic means. In particular, the general solution of (\ref{PU}) proves to be a
a linear combination of the four functions $\cos^3{\frac{t}{R}}$, $\sin^3{\frac{t}{R}}$, $\cos{\frac{t}{R}}\sin^2{\frac{t}{R}}$ and
$\sin{\frac{t}{R}}\cos^2{\frac{t}{R}}$ with arbitrary coefficients.

Although constants of the motion associated with other generators in the $l=\frac{3}{2}$ conformal Newton-Hooke algebra are redundant for integrating (\ref{PU}), let us display them in explicit form.
The invariance of the action functional (\ref{actionPU}) under time translations and spatial rotations yields
\bea
&&
\mathcal{H}=2x_i^{(3)}\dot{x}_i-{\ddot x}_i^2+\frac{10}{R^2}\dot{x}_i^2+\frac{9}{R^4} x_i^2,
\nonumber\\[2pt]
&&
\mathcal{M}_{ij}=x_i^{(3)} x_j-x_j^{(3)}  x_i+\frac{10}{R^2}(\dot{x}_i x_j-\dot{x}_j x_i)-{\ddot x}_i \dot{x}_j+{\ddot x}_j \dot{x}_i.
\nonumber
\eea
In order to find a realization of the dilatations and the special conformal transformations, it is important to observe that the vector generators corresponding to the transformations (\ref{transform}) fit in the general construction
of the $l$--conformal Newton-Hooke algebra in \cite{Galajinsky_2}. In particular, taking the missing generators as in \cite{Galajinsky_2}
\bea\label{genDK}
&&
D=\frac{1}{2} R \sin{(2t/R)} \frac{\partial}{\partial t}+\frac{3}{2}\cos{(2t/R)} x_i \frac{\partial}{\partial x_i},
\nonumber\\[2pt]
&&
K=R^2\sin^2{\frac{t}{R}}\frac{\partial}{\partial t}+\frac{3}{2}R \sin{(2t/R)} x_i\frac{\partial}{\partial x_i}
\eea
one can verify that the action functional (\ref{actionPU}) is invariant under the coordinate transformations generated by (\ref{genDK}) and
the corresponding constants of the motion read
\bea
&&
\mathcal{D}=-\frac{R}{4}\sin{\frac{2t}{R}} {\ddot x}_i^{2}+\frac{9}{2 R}\sin{\frac{2t}{R}}\dot{x}_i^{2}+\frac{R}{2}\sin{\frac{2t}{R}}\dot{x}_i x_i^{(3)}+\frac{1}{2}\cos{\frac{2t}{R}}\dot{x}_i {\ddot x}_i-\frac{3}{2}\cos{\frac{2t}{R}}x_i x_i^{(3)}-
\nonumber\\[2pt]
&&
\quad \quad
-\frac{9}{R^2}\cos{\frac{2t}{R}}x_i\dot{x}_i-\frac{3}{R} \sin{\frac{2t}{R}}x_i {\ddot x}_i- \frac{27}{4 R^3}\sin{\frac{2t}{R}}x_i^2,
\\[2pt]
&&
\mathcal{K}=-\frac{R^2}{2}\sin^2{\frac{t}{R}} {\ddot x}_i^2+\left(5\sin^2{\frac{t}{R}}-2\cos{\frac{2t}{R}}\right)\dot{x}_i^2+\frac{9}{R^2}\left(\frac{1}{2}\sin^2{\frac{t}{R}}+\cos{\frac{2t}{R}}\right)x_i^2+
\nonumber\\[2pt]
&&
\quad \quad
+ R^2\sin^2{\frac{t}{R}}\dot{x}_i x_i^{(3)}-\frac{9}{R} \sin{\frac{2t}{R}}x_i\dot{x}_i-\frac{3R}{2}\sin{\frac{2t}{R}}x_i x_i^{(3)}+3\cos{\frac{2t}{R}}x_i {\ddot x}_i+\frac{R}{2}\sin{\frac{2t}{R}}\dot{x}_i {\ddot x}_i.
\nonumber
\eea

Thus, we have demonstrated that for a particular choice of frequencies such that $\frac{\a^2}{\b^2}=\frac{10}{9}R^2$, the multi--dimensional Pais--Uhlenbeck oscillator enjoys the $l=\frac{3}{2}$--conformal
Newton-Hooke symmetry. It should be mentioned that a possibility to reveal such a symmetry in the Pais--Uhlenbeck oscillator (without mentioning the particular choice of the frequencies, however) was anticipated in \cite{Gomis}.

\vskip 0.5cm
\noindent
{\bf 5. Conclusion}

\vskip 0.5cm

To summarize, in this paper the method of nonlinear realizations and the technique previously developed in \cite{Galajinsky_1}
were used to construct a dynamical system without higher derivative terms, which holds invariant
under the $l$--conformal Newton-Hooke group.
A configuration space of the model involves coordinates, which
parametrize a particle moving in $d$ spatial dimensions and
a conformal mode. The latter gives rise to an effective external field. In accord with evolution of the conformal mode,
such a dynamical system describes a generalized multi--dimensional oscillator, which undergoes phases of accelerated and decelerated motion
in an ellipse.
We also considered higher derivative formulations and demonstrated that the multi--dimensional Pais--Uhlenbeck oscillator enjoys the $l=\frac{3}{2}$--conformal Newton-Hooke symmetry
for a particular choice of its frequencies.

Turning to possible further developments, it would be interesting to investigate whether the formulations considered in this work could be deformed by interaction terms in a way compatible with the $l$--conformal Newton-Hooke symmetry. It is interesting to construct a Lagrangian formulation for the models in Sect. 2 and Sect. 3. A possibility to realize the $l$--conformal Newton-Hooke symmetry in the multi--dimensional
Pais--Uhlenbeck oscillator for other choices of frequencies is worth studying as well.

\vskip 0.5cm
\noindent
{\bf Acknowledgements}\\

\noindent
This work was supported by the Dynasty Foundation,
RF Federal Program "Kadry" under the contracts 16.740.11.0469 and 14.B37.21.1298, MSE Program "Nauka" under the grant 1.604.2011,
and the LSS grant 224.2012.2.

\end{document}